**ORIGINAL PAPER**

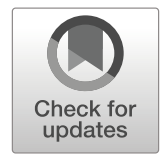

# Processes associated with ionic current rectification at a 2D-titanate nanosheet deposit on a microhole poly(ethylene terephthalate) substrate

Budi Riza Putra[1,2] · Christian Harito[3] · Dmitry V. Bavykin[3] · Frank C. Walsh[3] · Wulan Tri Wahyuni[2] · Jacob A. Boswell[1] · Adam M. Squires[1] · Julien M. F. Schmitt[1] · Marcelo Alves Da Silva[1] · Karen J. Edler[1] · Philip J. Fletcher[4] · Anne E. Gesell[4] · Frank Marken[1]

**Abstract**
Films of titanate nanosheets (approx. 1.8-nm layer thickness and 200-nm size) having a lamellar structure can form electrolyte-filled semi-permeable channels containing tetrabutylammonium cations. By evaporation of a colloidal solution, persistent deposits are readily formed with approx. 10-μm thickness on a 6-μm-thick poly(ethylene-terephthalate) (PET) substrate with a 20-μm diameter microhole. When immersed in aqueous solution, the titanate nanosheets exhibit a p.z.c. of − 37 mV, consistent with the formation of a cation conducting (semi-permeable) deposit. With a sufficiently low ionic strength in the aqueous electrolyte, ionic current rectification is observed (cationic diode behaviour). Currents can be dissected into (i) electrolyte cation transport, (ii) electrolyte anion transport and (iii) water heterolysis causing additional proton transport. For all types of electrolyte cations, a water heterolysis mechanism is observed. For $Ca^{2+}$ and $Mg^{2+}$ions, water heterolysis causes ion current blocking, presumably due to localised hydroxide-induced precipitation processes. Aqueous $NBu_4^+$ is shown to 'invert' the diode effect (from cationic to anionic diode). Potential for applications in desalination and/or ion sensing are discussed.

**Keywords** Voltammetry · Ion valve · Ionic logic · Sensing · Nanostructure · Iontronics

## Introduction

Ion binding and transport through microporous and mesoporous materials is of considerable importance in water purification and in membrane science [1, 2]. Recent interest has arisen in particular in the porosity of new types of lamellar nanostructures (in particular graphene oxide [3, 4]), which are linked to ion transport [5] and water transport [6] phenomena, and the desire to exploit new types of lamellar 2D-materials more widely [7]. There are now ranges of novel materials available based, for example on 2D-nanosheets of modified carbons [8], phosphorous [9], oxides [10], chalcogenides [11] and mixed oxides such as $BiVO_4$ [12] and MXenes [13]. 2D-titanates have been of considerable interest and lamellar titanate deposits have been shown to be electrochemically and photo-electrochemically active depending on the pH and intercalated guest species [14]. In this report, we focus on titanate nanosheet materials that have been developed in pioneering work by Sasaki and coworkers [15, 16].

Titanate nanosheet materials are produced in two steps [17]; $TiO_2$ reacts with caesium salts to give an intermediate solid lepidocrocite-type product, which is exfoliated into aqueous solution containing alkylammonium cations (here tetrabutylammonium, [18]) to give a stable colloidal solution

✉ Frank Marken
f.marken@bath.ac.uk

[1] Department of Chemistry, University of Bath, Claverton Down, Bath BA2 7AY, UK

[2] Department of Chemistry, Faculty of Mathematics and Natural Sciences, Bogor Agricultural University, Bogor, West Java, Indonesia

[3] Energy Technology Research Group, Faculty of Engineering and the Environment, University of Southampton, Southampton SO17 1BJ, UK

[4] Materials and Chemical Characterisation Facility (MCC), University of Bath, Claverton Down, Bath BA2 7AY, UK

of negatively charged titanate nanosheets. The surface charge of the colloid is associated with a point of zero charge (p.z.c.) at approximately pH 4 [19]. The colloidal solutions contain single unit cell thick titanate layers, which are typically 200 nm in diameter [20]. The deposits of this colloid when dried up (without heating) give a lamellar gel with typically 1.7 to 1.8 nm unit cell spacing (based on the X-ray diffraction pattern, [21]). This is consistent with an approximately 1.2- to 1.3-nm-thick titanate unit cell layer sandwiched between electrolyte layers of typically 0.5 nm. At elevated temperature, these types of gels can collapse back to the anatase crystal form but, at room temperature, these lamellar nanostructures persist.

Applications of titanate nanosheets have been proposed as component in batteries [22], in heterogeneous catalysis [23], in photocatalysis [24] and as functional filler in polymer blends [25]. Recently, we have demonstrated that a film deposit of titanate nanosheet material on a glassy carbon electrode allows ferroceneboronic acid to be immobilised and binding of fructose within the nano-lamellar space to be detected [20]. Further study has shown that the titanate deposits behave more similar to organic media (and not to hydrophilic oxides) with the ability to bind hydrophobic organic molecules (e.g. anthraquinone) from cyclopentanone and to allow redox conversion of these immobilised inter-lamellar redox systems [26]. The ability to host hydrophobic guests has been attributed to the presence of tetrabutylammonium cations within the electrolyte layer.

In order to investigate ion transport through the lamellar titanate nanosheets, deposits of the titanate are studied here on a poly(ethylene-terephthalate) substrate (PET with 6-μm thickness) with a microhole (20-μm diameter). The experimental arrangement is very similar to that introduced by Girault et al. [27] for the study of liquid|liquid micro-interfaces. Recently, other types of ionomer systems (such as reconstituted cellulose [28], Nafion™ [29], the commercial ionomer Fumasep™ [30], graphene oxide [31] and a polymer of intrinsic microporosity [32]) have been investigated on this type of microhole substrate in order to reveal the mechanisms/characteristics for ionic current rectification (or 'ionic diode' phenomena). The ionic diode or ionic current rectification effect requires a semi-permeable material. This type of effect allows cation transport (cationic diodes for cation conductors) and anion transport (anionic diodes for anion conductors) to be distinguished. New ideas for ionic current rectifier applications have been proposed, for example for water purification [33] or for ionic diode sensing [34].

The ionic diode effect caused in asymmetrically coated microholes can be explained (at least to a first approximation) based on a combination of ion conductivity and concentration polarisation phenomena [35]. Figure 1 provides a schematic summary of the key processes. With a symmetric ionomer deposit on both sides of the microhole, transport of ions from the left or from the right compartment remains the same and a simple resistance is observed rather than rectifier behaviour. However, with semi-permeable ionomer deposited only on one side (here, always on the side of the working electrode in a classic four-electrode measurement cell [36]), a distinct change in current occurs with an 'open' potential domain (due to accumulation of electrolyte in the microhole region; current flows) and a 'closed' potential domain (due to depletion of electrolyte in the microhole region; current is blocked). This is illustrated in Fig. 1 for the case of a cationic diode.

Ionic diode phenomena are associated with processes that occur at different length scales due to potential dependent compositional changes. There are different types of mechanisms, for example based on double-layer changes within nanopores, based on diffusion-migration layer compositional changes and due to interfacial precipitation or blocking. The diffusion-migration layer–based processes in microhole diodes (as described in Fig. 1) can be contrasted with processes reported in work on nanochannel diodes [37], nano-cones [38] and on electrolytic microfluidic/nanofluidic channel diodes [39, 40]. Other types of ionic diodes have been proposed based on gel-gel interfaces [41]. Ionic circuits have been proposed based on polymer gels [42, 43]. Ionic diode switching effects have been demonstrated also due to interfacial precipitation [44] and due to pH gradients [45]. The term 'iontronics' has been coined by Chun and Chung [46] to emphasise the excitement associated with developments functionality, such as that in ionic amplifiers [47], transistors [48] or flip-flops [36]. More fundamentally, ionic rectifiers under AC-excitation can be considered as 'ion pumps'. A device based on coupling cationic diodes and anionic diodes has been proposed for desalination applications [33]. Selectivity in ionic diode behaviour is desirable in order to broaden the possible range of applications. In particular, selectivity for higher valent cations such as $Mg^{2+}$ and $Ca^{2+}$ could be very useful for water treatment applications.

In this report, ion transport in the inter-lamellar space of titanate nanosheets is investigated. The semi-permeable nature of titanate nanosheet deposits (immersed in electrolyte with not too high ionic strength) allows cation transport and therefore causes cationic diode effects. This is studied here for a wider range of electrolytes and conditions. Inversion to anionic diode behaviour is observed in the presence of tetrabutylammonium cations.

# Experimental

## Chemical reagents

Titanate nanosheet material was synthesised as described previously by Sasaki et al. [21] and by Harito et al. [49, 50]. Hydrochloric acid, sodium chloride, potassium chloride, lithium chloride, ammonium chloride, tetrabutylammonium chloride, magnesium chloride, calcium chloride, sodium

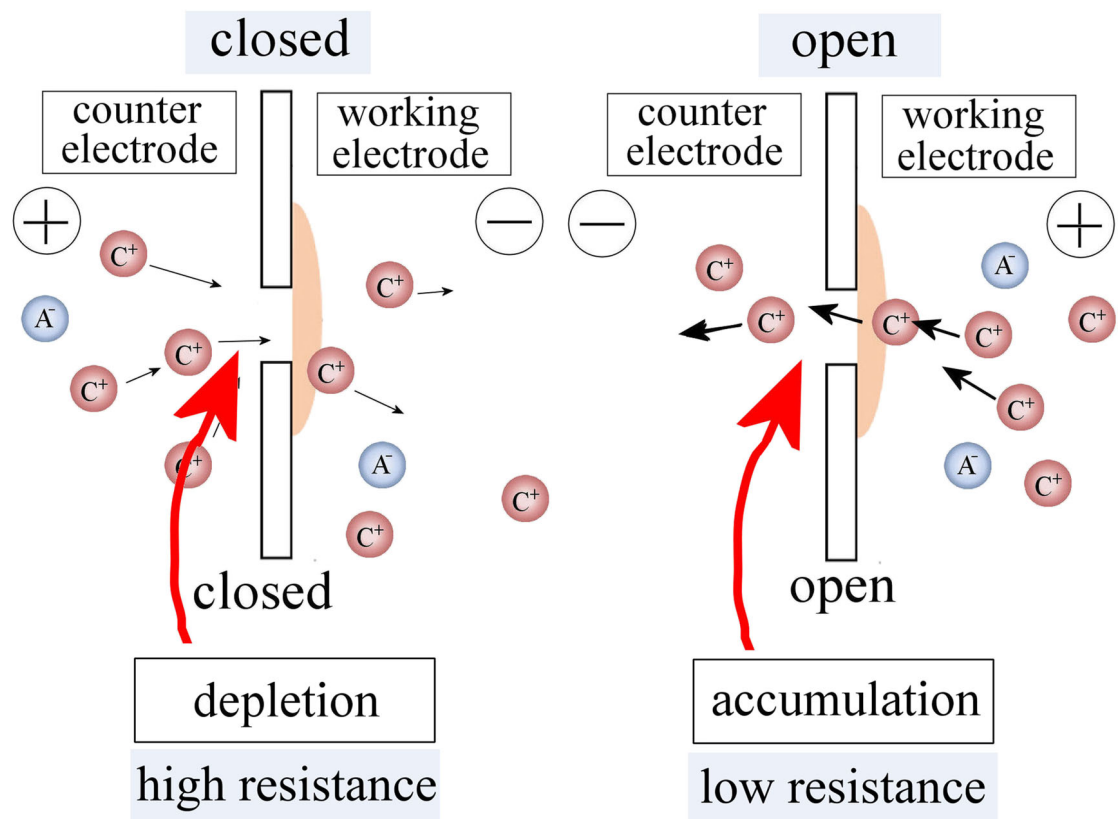

**Fig. 1** Schematic depiction of the case of a semi-permeable cation conductor causing a 'cationic diode' effect. In the closed state, the flow of cations is limited by electrolyte depletion within the cylindrical section of the PET film. In the open state, electrolyte accumulation within the cylindrical section occurs due to oversupply of cations through the semi-permeable cation conductor

hydroxide, sodium nitrate, sodium perchlorate and sodium sulphate were obtained in analytical grade from Sigma-Aldrich or Fischer Scientific and used without further purification. Phosphate buffer saline (PBS) was prepared with $NaH_2PO_4$, $Na_2HPO_4$ and NaCl and adjusted to pH 7. Solutions were prepared under ambient condition in volumetric flasks with ultrapure water with resistivity of 18.2 MΩ cm (at 22 °C) from an ELGA Purelab Classic System.

### Instrumentation

Electrochemical data (voltammetry and chronoamperometry) were recorded at $T = 20 \pm 2$ °C on a potentiostat system (Ivium Compactstat, The Netherlands). A classic four-electrode membrane electrochemical cell was used. The membrane separates two tubular half-cells (15-mm diameter, see Fig. 2a), one with Pt wire working and saturated calomel (SCE) sense electrode and the other with SCE reference electrode and Pt wire counter electrode. In electrochemical measurements, the working electrode was always located on the side of titanate nanosheet films. Fluorescence imaging was performed on a Carl Zeiss Confocal Scanning Microscope. For fluorescence analysis, rhodamine B was mixed with titanate nanosheet colloidal solution, which was then applied to the PET films (see below).

### Procedures

The zeta potential for the colloidal titanate solution was measured on a Zetasizer Nano ZS (Malvern Instruments Ltd., Malvern, UK) confirmed an average of − 32.5 mV and an average size of 216 nm [26]. In order to form films of titanate nanosheets on PET substrates (20-μm diameter hole in 6-μm-thick PET from Laser Micromaching Ltd., Birmingham, UK), 10 μL aqueous titanate nanosheet colloidal solution (2.56 g $L^{-1}$) was applied to a PET film on a glass substrate (the glass was pre-coated with a thin layer of 1% agarose gel by solution casting to stop titanate nanosheet material penetrating through the PET microhole). A volume of 10 μL titanate nanosheet colloidal solution was applied to the surface over PET to give a 1-$cm^2$ film coating, which after drying produced a thin uniform coating. For the formation of symmetric deposits on both sides of the PET film, the deposition was repeated on the back side of the PET. In order to image the $TiO_2$ films, stacked fluorescence images were obtained for a symmetric (double-sided) and for the asymmetric (one-sided) deposits.

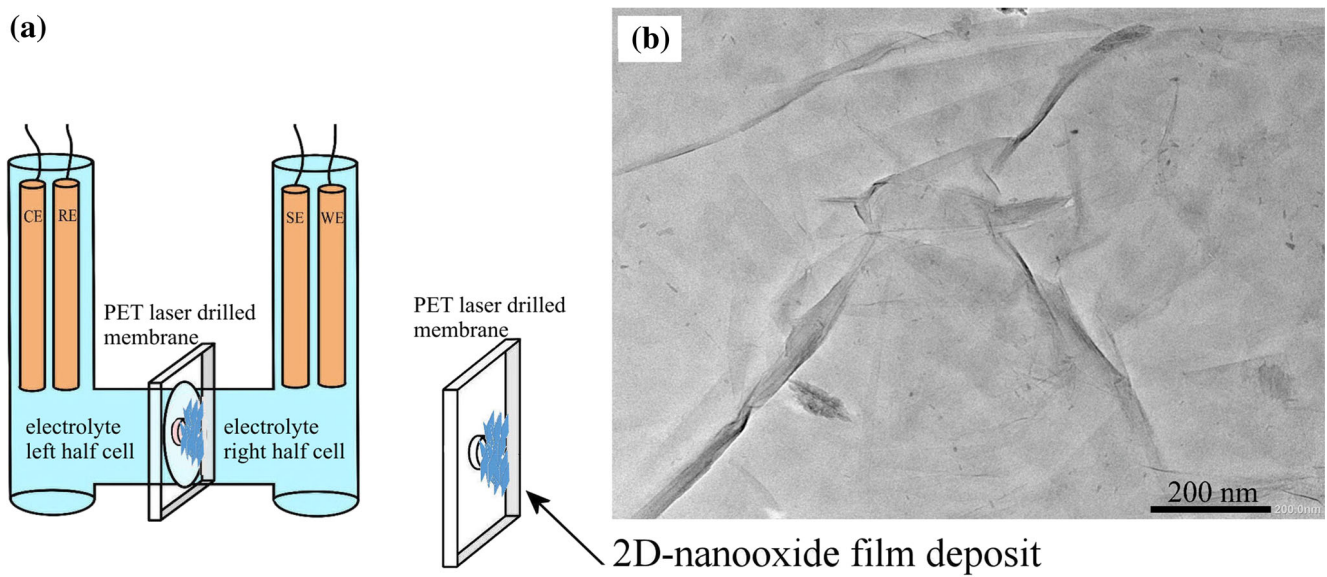

**Fig. 2** a Schematic description of the experiment with four-electrode control (WE, working electrode; SE, sense electrode; CE, counter electrode; RE, reference electrode) of the applied potential and TiO$_2$ nanosheet deposit on the side of the working electrode. b A typical TEM image of the titanate nanosheet material is shown

### Titanate nanosheet film characterisation

Fluorescence microscopy with a rhodamine B stain reveals the presence of a uniform ~ 5- to 10-μm-thick film of titanate nanosheets coated over a 20-μm diameter microhole in a PET substrate. Figure 3 shows a multi-stack fluorescence microscopy image as top view (with the focus on the PET layer) and as side view. For the symmetric diode (Fig. 3a), it can be seen that the 20-μm microhole was partially filled with titanate nanosheet material. It can be clearly observed that there

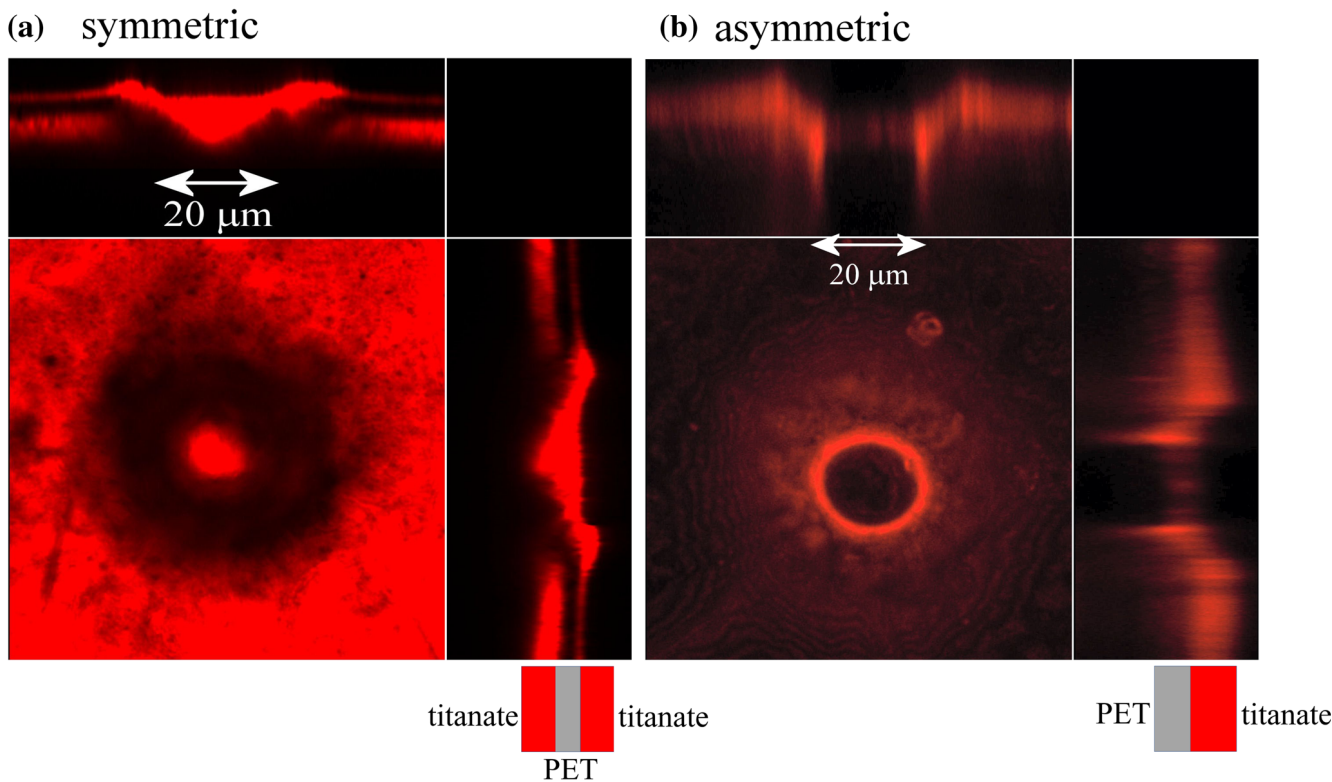

**Fig. 3** Fluorescence microscopy image stacks based on rhodamine B embedded into TiO$_2$ nanosheet deposits a symmetric (applied from both sides) and b asymmetric (applied form only one side) to the PET substrate with 20-μm diameter microhole

are two layers of titanate nanosheets on opposite sides of the PET. However, for the asymmetric diode (Fig. 3b), the layer of titanate nanosheet material is seen on one side (ca. 5–10 μm thick). The extent of the deposit is not clearly resolved due to light reflection effects.

## Results and discussion

### Ion transport through lamellar titanate nanosheet deposits: Na$^+$ transport

Figure 4 a shows typical cyclic voltammetry data over a potential range of + 4 to − 4 V applied to the nanosheet coated or uncoated PET membrane immersed in 10 mM NaCl on both sides. Three cases are shown: (i) a PET film with an empty 20-μm diameter hole, (ii) a PET film with titanate nanosheets deposited symmetrically on both sides and (iii) a PET film with titanate nanosheets deposited asymmetrically only on one side. For the empty microhole, the currents are dominated by the specific resistivity (giving a constant resistance) of the aqueous NaCl electrolyte solution filling the microhole [51]. With titanate nanosheets applied symmetrically on both sides, the current decreases for both negative and positive applied potentials due to the titanate restricting the flow of ions. For the asymmetric case, current rectification behaviour is observed with a higher current generated at positive applied potentials and a lower current generated at negative applied potentials. Figure 1 illustrates the corresponding cases of an open cationic diode (positive applied potential) and a closed cationic diode (negative applied potential) for the case of negatively charged titanate nanosheet material. Next, rectification phenomena are investigated with different concentrations of aqueous NaCl electrolyte solution.

Figure 4 b shows cyclic voltammetry data for asymmetrically deposited titanate nanosheets. Rectified currents flow with positive applied potentials and therefore cation conduction must be possible through the titanate deposit. With the increasing concentration of NaCl from 1 to 1000 mM, a well-defined steady-state voltammetric response is observed. For experiments at lower NaCl concentration, the current in the negative potential range is lower compared to the current in the positive potential range. This is consistent with cationic diode behaviour as seen in cases such as Nafion™ [29] and Fumasep™ FKS-30 ionomer [30]. The effect of NaCl electrolyte concentration on the currents and on the switching time for the cationic diode is shown more clearly in chronoamperometry data in Fig. 4 c. From these data, the rectification ratio can be calculated (obtained by dividing the absolute currents at + 1 V and at − 1 V; see inset in Fig. 4 b). An increase in NaCl concentration causes an increase in both the current in the open state and the current in the closed state. An optimum in rectification effect is observed in the range

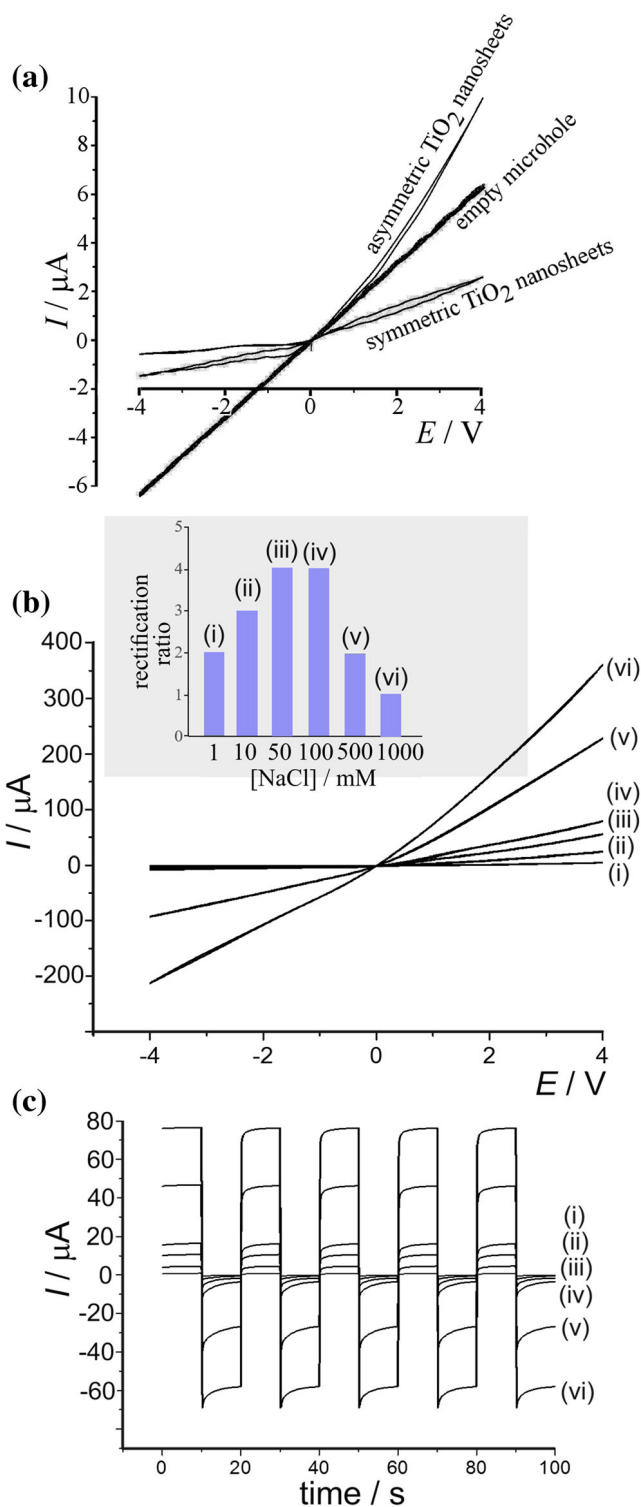

**Fig. 4** **a** Cyclic voltammograms (scan rate 50 mV s$^{-1}$) for (i) an empty microhole, (ii) a symmetric titanate deposit and (iii) an asymmetric titanate deposit (microhole diameter 20 μm) immersed in 10 mM NaCl. **b** Data for asymmetrically deposited titanate nanosheets showing cyclic voltammograms (scan rate 25 mV s$^{-1}$) in aqueous NaCl and **c** chronoamperometry responses (stepping from + 1 to − 1 V) immersed in aqueous NaCl with a concentration of (i) 1, (ii) 10, (iii) 50, (iv) 100, (v) 500, (vi) 1000 mM NaCl on both sides. Inset: rectification ratios from chronoamperometry data at ± 1 V

from 50 to 100 mM NaCl solution. The decrease in rectification ratio when going to higher ionic strength can be explained due to loss of semi-permeability and onset of anion transport through the titanate nanosheet film.

## Ion transport through lamellar titanate nanosheet deposits: electrolyte cations versus anions

Next, the effects of electrolyte cations and anions on titanate nanosheet transport are investigated in order to better understand cationic diode processes. Figures 5 a–b show a summary of data from cyclic voltammogram, chronoamperometry and rectification ratio bar plots for aqueous 10 mM HCl, NaCl, KCl, LiCl, $NH_4Cl$, $MgCl_2$ and $CaCl_2$. All salt systems show some conductivity, but there are distinct changes depending on the type of cation. Currents for the open diode are considerably higher in the presence of monovalent cations such as $H^+$, $Li^+$, $Na^+$, $K^+$ and $NH_4^+$ compared to those for divalent cations such as $Mg^{2+}$ and $Ca^{2+}$. This could be consistent with a higher mobility for monovalent cations relative to divalent cations. However, the ability of cations to bind/transfer is affected by water heterolysis and will be investigated in more detail below. Among the monovalent cations, protons gave the highest currents and rectification ratio (Fig. 5a–b).

Next, the effects of electrolyte anions are investigated. Figures 5 c–d show a summary of data from cyclic voltammogram, chronoamperometry and a rectification ratio bar plot for aqueous 10 mM NaOH, $NaNO_3$, NaCl, $NaClO_4$, $Na_2SO_4$ and PBS pH 7. For all types of anions, similar levels of rectification are observed. Currents for the open diode (in the positive potential range) become significantly higher for higher valent anions ($SO_4^{2-}$ and $PO_4^{3-}$), but effects of the type of anion on the rectification ratio data seem insignificant. Clearly, anions at a sufficiently high concentration can enter into the inter-lamellar space and significantly change the rate of transport. This suggests that there could be always an underlying contribution from anion transport, even when the cation transport appears dominant. Sulphate and phosphate are likely to more strongly collapse the internal double layer and thereby to open up the nanochannels to both cation and anion transport. This can give rise to a higher ion current and less semi-permeability. Chronoamperometry data show relatively complex transients (rising and falling) with transient features mainly taking place in the first 0.5 s of the data. It is likely that during this period, not only concentration gradients in the solution in the vicinity of the microhole develop, but also that concentration gradients within the nanosheet material change. The current traces measured during chronoamperometry may contain a component of 'adsorption currents' that are associated with compositional changes in the inter-lamellar space. Although electrolyte anion effects are clearly significant, it is currently difficult to dissect these and to provide a more detailed assessment of the nature of these effects.

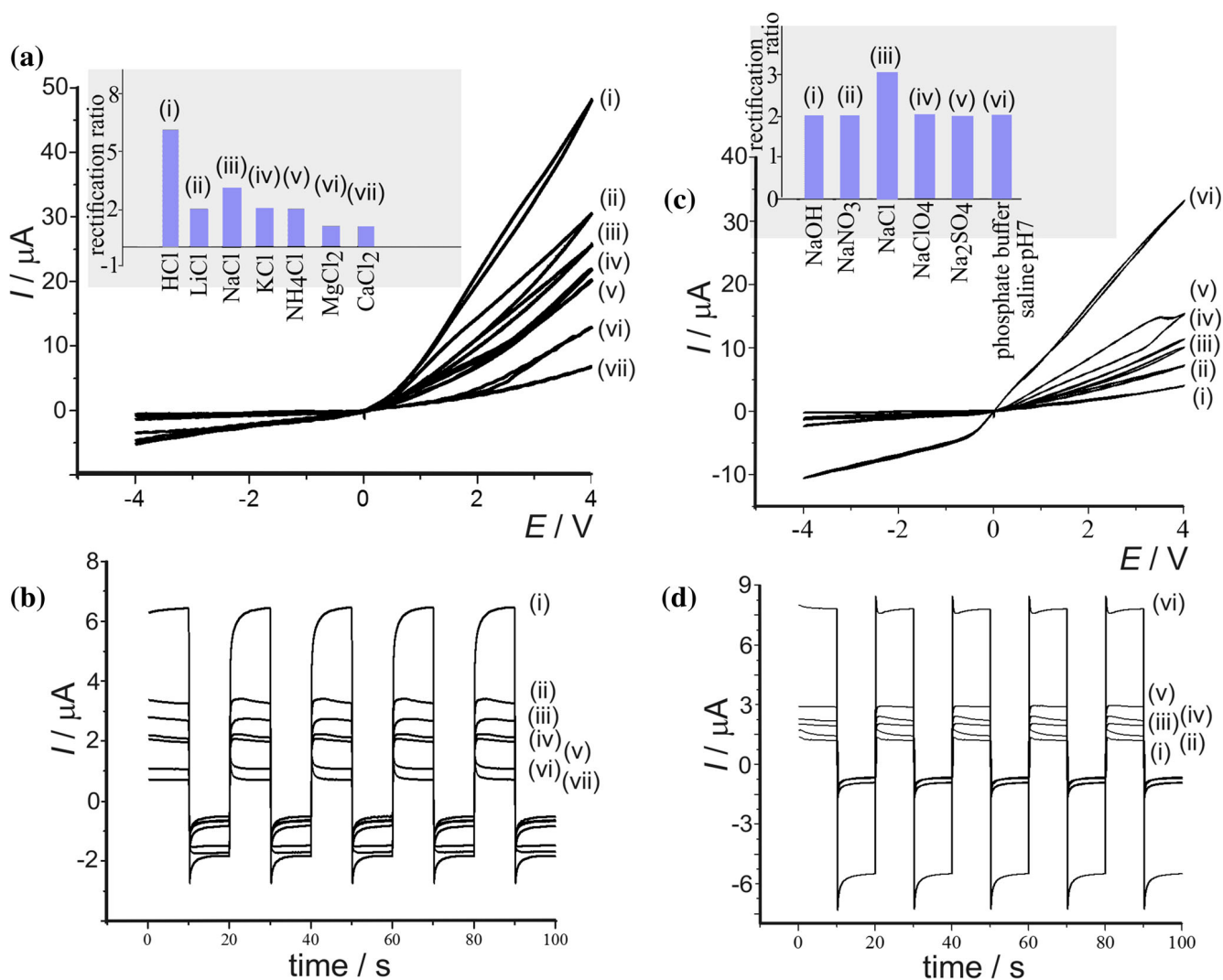

**Fig. 5** Data for asymmetric titanate nanosheet deposits. **a** Cyclic voltammograms (scan rate 25 mV s$^{-1}$), **b** chronoamperometry data (stepping from + 1 to − 1 V), immersed in aqueous (i) HCl, (ii) LiCl, (iii) NaCl, (iv) KCl, (v) $NH_4Cl$, (vi) $MgCl_2$ and (vii) $CaCl_2$ (10 mM on both sides). **c** Cyclic voltammograms (scan rate 25 mV s$^{-1}$), **d** chronoamperometry data (stepping from + 1 to − 1 V), for films immersed in aqueous (i) NaOH, (ii) $NaNO_3$, (iii) NaCl, (iv) $NaClO_4$, (v) $Na_2SO_4$ and (vi) phosphate buffer saline pH 7 (all 10 mM on both sides). Insets show bar graphs for current rectification (see text)

## Ion transport through lamellar titanate nanosheet deposits: competing cation and proton transport

The observation of current through the titanate nanosheet deposit can be associated generally with three types of processes based on (i) flow of cations, (ii) flow of anions or (iii) the potential-driven water heterolysis process leading to the formation of mobile protons and mobile hydroxide (with a resulting net pH gradient). The latter process has been reported for biological [52] and for bipolar membrane processes [53]. Catalysts for heterolytic water splitting such as graphene oxide [54] have been proposed to considerably enhance the potential-driven proton and hydroxide formation. In measurements as those reported in Fig. 5, it is not immediately clear whether water heterolysis can contribute to or even dominate the observed currents.

One way of directly monitoring the simultaneous formation of protons and hydroxide is by placing a secondary working electrode (WE2) into the two-compartment cell. Figure 6 shows a schematic drawing with WE2 place in the vicinity of the membrane and in the left counter electrode compartment. Initially, a cyclic voltammetry experiment was performed (with the three electrodes in the left compartment) to observe the onset potential for hydrogen evolution. In Fig. 6 a, data for 10 mM HCl are shown and the inset shows the onset at − 0.4 V vs. SCE (indicated by an arrow). Next, the five-electrode membrane cell is operated in cyclic voltammetry mode with the collector (WE2) potential fixed at the onset potential. Transport of protons associated with the open diode (at positive applied potentials) causes an increase in proton concentration at the location of WE2 and thereby leads to a current.

Figure 6 a shows both the generator current (WE1, membrane) and the collector current (WE2). The magnitude of the collector current approaches that of the generator current (with the opposite sign) in the case of a pure proton current. Therefore, the data for the proton transport for 10 mM HCl in Fig. 6 a provide proof for the methodology to work. Data in Fig. 6 b are shown for 10 mM LiCl. Perhaps surprisingly, again a significant proton transport is observed with the collector current almost mirroring the generator current. This suggests that water heterolysis at titanate nanosheets is important and that the transport of $Li^+$ is likely to be accompanied by proton transport. Results for NaCl (Fig. 6c), KCl (Fig. 6d) and $NH_4Cl$ (Fig. 6e) are very similar, and also suggest proton formation by heterolysis and transport through the titanate film.

Data observed for $MgCl_2$ (Fig. 6f) and for $CaCl_2$ (Fig. 6g) are different not only with much lower proton fluxes, but also much lower generator currents. The shape of the voltammetric signal for $Ca^{2+}$ is more complex and indicative of some blocking of the diode as the potential is scanned into the positive potential range. For both $Mg^{2+}$ and $Ca^{2+}$, small transient peaks are seen on the negative going potential scan, which could be associated with an 'unblocking' process. These observations are consistent with a blocking effect for $Mg^{2+}$ and $Ca^{2+}$ caused by a localised pH changes at the titanate film surface. Water heterolysis occurs at the titanate nanosheets, and this then causes locally the formation of hydroxide which then in combination with $Mg^{2+}$ or $Ca^{2+}$ causes blocking of the titanate nanochannels. However, the importance of this effect and the relative importance of proton versus electrolyte cation transport through the titanate film are difficult to quantify from these data. Further evidence for these processes comes from direct pH measurements.

In Fig. 7, a schematic drawing illustrates how two pH probes (glass membrane-based, battery controlled) are incorporated into the four-electrode electrochemical cell. The pH probes give a reading for the local pH (as an average) close to the PET film on the left and on the right. The photograph in Fig. 7 a shows the system before starting the experiment. With 10 mM NaCl in the left and right compartments, the pH is close to neutral. When operating the cationic diode at + 4 V (open), an average current of approximately 23 μA flows and the pH value start to divert. After 10 min of operation (see Fig. 7b) the pH in the left compartment reaches 6.56 (more acidic) and in the right compartment 6.91 (less acidic). Figure 7 c–d summarise pH data for experiments in 1 mM NaCl and in 10 mM NaCl. For a concentration of 100 mM NaCl, pH changes were less significant. It is possible to relate these observations to the water heterolysis process.

The effect of protonic current flow through the diode can be estimated in terms of an upper limit for the pH changes in the right/left compartment when assuming a fixed volume $V$ of ca. 0.01 dm$^3$ (an estimated volume where pH changes occur close to the membrane surface). Eqs. (1) and (2) express the shift in pH (from the initial value pH$_{initial}$) in the right compartment (where protons are lost) and in the left compartment (where protons are gained).

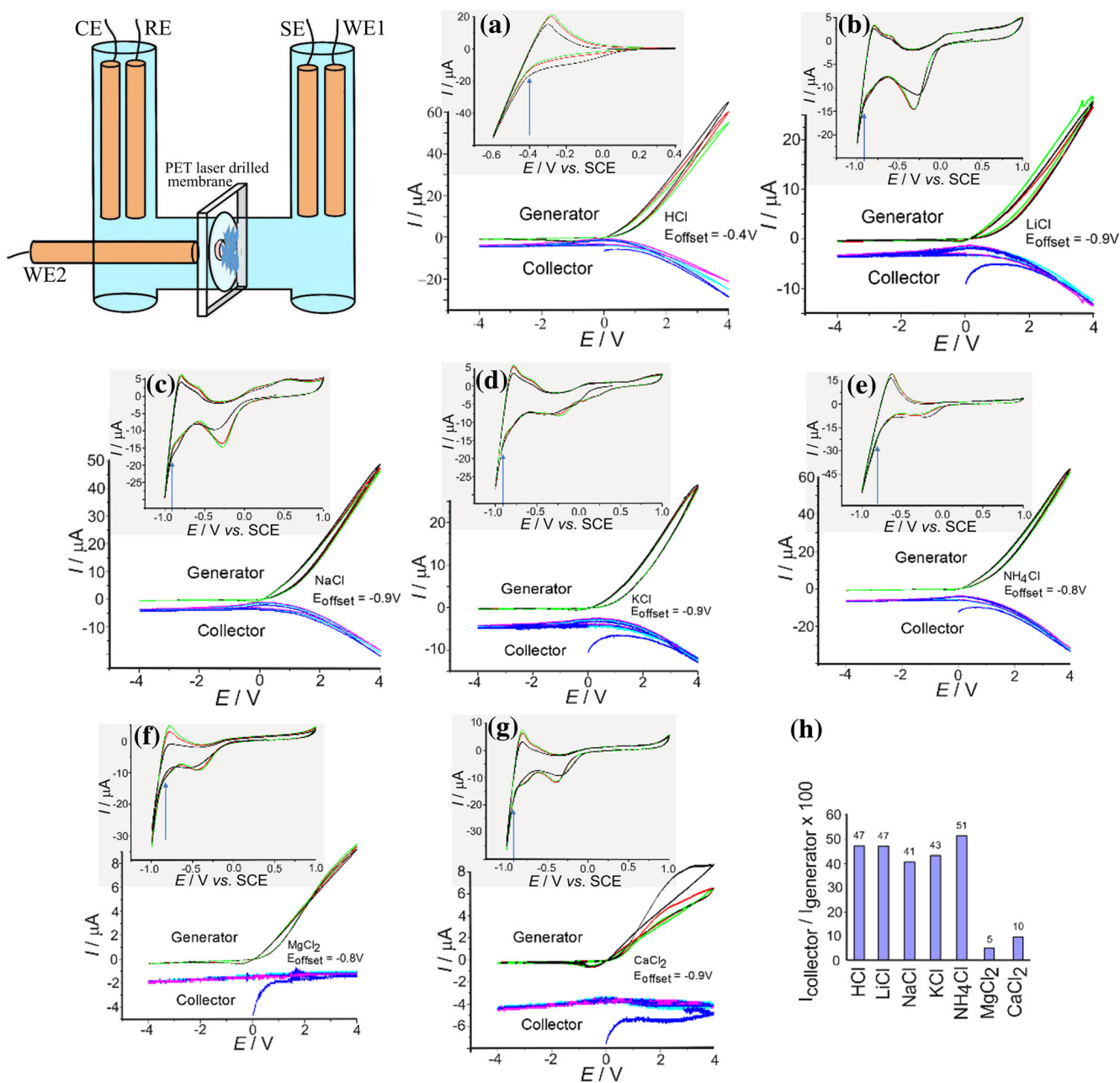

**Fig. 6** Schematic drawing showing the five-electrode cell with additional (WE2, 3-mm diameter platinum disk) collector electrode for sensing proton flux. Data sets for **a** 10 mM HCl, **b** 10 mM LiCl, **c** 10 mM NaCl, **d** 10 mM KCl, **e** 10 mM NH$_4$Cl, **f** 10 mM MgCl$_2$ and **g** 10 mM CaCl$_2$ including the cyclic voltammogram (scan rate 50 mV s$^{-1}$) at WE2 showing onset of hydrogen evolution (blue arrow) and the generator-collector voltammogram (three consecutive potential cycles, scan rate 50 mV s$^{-1}$). **h** Bar graph comparing the ratio of collector current and generator current (at + 4 V; in %) for different electrolyte media

$$\text{pH}_{\text{left}} = -\log_{10}\left(10^{-\text{pH}_{\text{initial}}} + It/FV\right) \quad (1)$$

$$\text{pH}_{\text{right}} = 14 + \log_{10}\left(10^{-(14-\text{pH}_{\text{initial}})} + It/FV\right) \quad (2)$$

Parameters in these equations are the time-average absolute current $I$, the time $t = 600$ s and the Faraday constant $F = 96,487$ C mol$^{-1}$. For the case of a 1 mM NaCl electrolyte (see Fig. 7c), the average current at an applied voltage of 4 V was 10 μA over 600 s, which (with pH$_{initial}$ = 6.39) translates to pH$_{left}$ = 5.2 and pH$_{right}$ = 8.8. Comparison to the experimental results in Fig. 7 c shows that the predicted trend is indeed

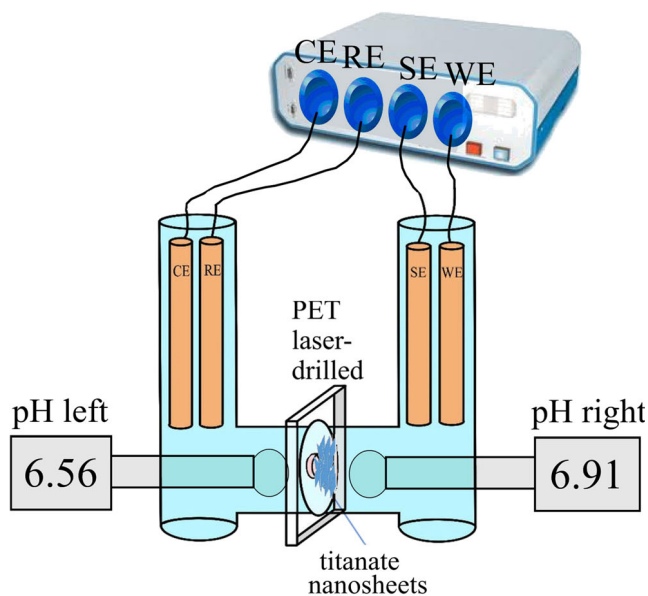

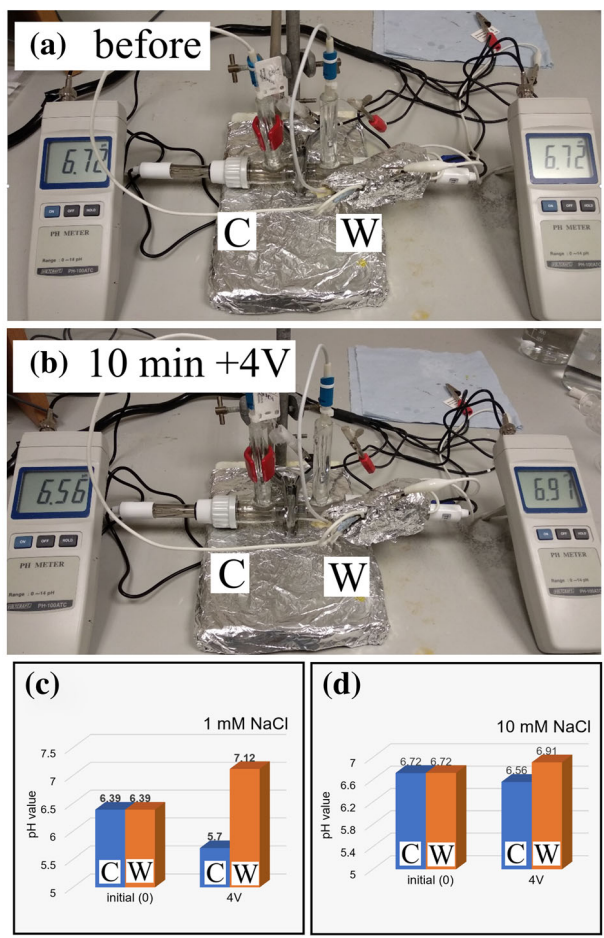

**Fig. 7** Illustration of the experimental system. **a** Photograph showing the experimental cell with the left compartment (counter electrode and reference electrode) and the right compartment (working electrode and sense electrode). In both compartments, an additional battery-driven pH sensor is added to probe the pH close to the PET film in the middle. **b** After 10 min + 4 V and 23 μA current the pH has become more acidic on the left and more alkaline on the right. Bar graphs for the effect at 10 min + 4 V for **c** 1 mM NaCl (10 μA) and **d** 10 mM NaCl (23 μA)

observed. For a concentration of 10 mM NaCl (Fig. 7d), the observed average current was 23 μA at 4 V applied voltage. This gives $pH_{left}$ = 4.8 and $pH_{right}$ = 9.2. Experimental data suggest a less strong change in pH and therefore less competition between $Na^+$ transport and water heterolysis at the higher NaCl concentration. Additional experiments for 100 mM NaCl (not shown) suggest even lower pH drift even at higher currents. Water heterolysis appears to be strongly electrolyte concentration dependent and probably insignificant at electrolyte concentrations of 10 mM or higher. Although conclusive in terms of providing direct evidence for the water heterolysis process, the data provide only qualitative insights and further work will be required to provide a quantitative measure of the water heterolysis process under these conditions.

### Ion transport through lamellar titanate nanosheet deposits: interference of $NBu_4^+$ with proton transport

The presence of $NBu_4^+$ cation in the original titanate host material is due to the exfoliation process requiring a bulky tetraalkylammonium reagent. When depositing titanate nanosheets onto the PET film, $NBu_4^+$ cations are present and they define the ability of the titanate lamellae to bind to host molecules [24] and to conduct ions. It is therefore important to also investigate the effect of $NBu_4^+$ cations on the ionic current rectification and on the ion conductivity. Figure 8 a shows membrane voltammetry data for 10 mM HCl electrolyte initially in both sides of the two-compartment cell. The typical cationic diode response is observed (Fig. 8a(i)). When adding $NBu_4Cl$ into the right compartment, this can equilibrate with the titanate nanosheet deposit and change the ability of the lamellar structure to conduct ions. Data in Fig. 8 a demonstrate a systematic decrease in ion current (and in rectification ratio) as the concentration of $NBu_4Cl$ increases. Ultimately, when adding 50 mM $NBu_4Cl$, the ionic diode is inverted. The switch of the open diode state from positive potential to negative potential is a strong indication for a change from a cationic diode to an anionic diode. This is also confirmed in chronoamperometry data in Fig. 8 b. The current transient shows a rising current when opening the diode. In the presence of 50 mM $NBu_4Cl$, the behaviour switches and the rising current transient is observed in the negative potential domain.

It is possible to envisage the processes responsible for the ion transport in the lamellar structure based on the equilibration of competing cation in the titanate double layer. Both protons and $NBu_4^+$ can be bound (Fig. 8c). In the absence of $NBu_4^+$ vacancies exist for protons (or other cations) to bind and to conduct through the titanate deposit. However, when adding $NBu_4^+$ cations these bind and block the transport of other cations (protons). For 10 mM $NBu_4^+$ cations both the current in the open state and the rectification ratio half. When adding more $NBu_4Cl$, it is possible to 'invert' the diode behaviour. This can be explained by excess of $NBu_4Cl$ binding

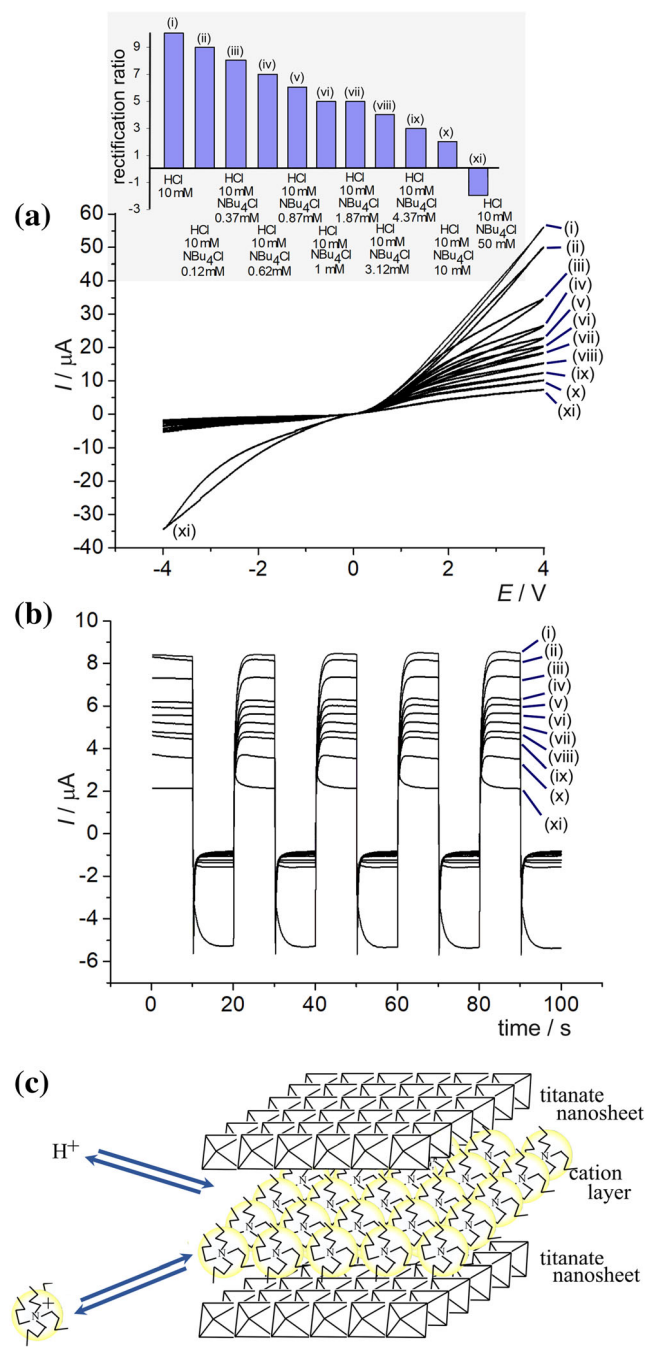

**Fig. 8** **a** Asymmetric TiO$_2$ nanosheet data for cyclic voltammograms (scan rate 25 mV s$^{-1}$) in aqueous electrolyte (for 10 mM HCl on the left and (i) 10 mM HCl, (ii) 10 mM HCl + 0.12 mM NBu$_4$Cl, (iii) 10 mM HCl + 0.37 mM NBu$_4$Cl, (iv) 10 mM HCl + 0.62 mM NBu$_4$Cl, (v) 10 mM HCl + 0.87 mM NBu$_4$Cl, (vi) 10 mM HCl + 1 mM NBu$_4$Cl, (vii) 10 mM HCl + 1.87 mM NBu$_4$Cl, (viii) 10 mM HCl + 3.12 mM NBu$_4$Cl, (ix) 10 mM HCl + 4.37 mM NBu$_4$Cl, (x) 10 mM HCl + 10 mM NBu$_4$Cl, (xi) 10 mM HCl + 50 mM NBu$_4$Cl, on the right). **b** Chronoamperometry data (stepping from + 1 to − 1 V) immersed in aqueous electrolyte. Inset: rectification ratio calculation from chronoamperometry data at ± 1 V. **c** Schematic description of the competition of protons and tetrabutylammonium cations

into the inter-lamellar space and a net anion transport mechanism (due to mobile chloride) being activated.

More work will be needed to further exploit this switch from cationic to anionic current rectification.

## Summary and conclusions

It has been shown that asymmetry in the deposition of a titanate nanosheet film can be used to induce current rectification phenomena or ionic diode effects, as long as the semi-permeable character due to cation transport is maintained. In this case, a 'cationic diode' has been produced with rectification effects observed for many types of cations. A cation vacancy model is proposed based on NBu$_4^+$ cations occupying part of the inter-lamellar space with vacancies for other types of cations to bind. The excess of negative charges on the titanate allow cation binding and transport.

The rectification ratio for NaCl electrolyte considerably decreased when concentration of aqueous NaCl (or the ionic strength) was increased on the side of titanate nanosheet film (consistent with anion uptake into the lamellar space and anion transport in addition to cation transport at higher ionic strength). Currents for the open diode are shown to be associated with additional water hydrolysis and proton transport rather than pure Li$^+$, Na$^+$, K$^+$, NH$_4^+$, Mg$^{2+}$ and Ca$^{2+}$ transport. In the case of Mg$^{2+}$ and Ca$^{2+}$, this resulted in blocking of transport presumably due to precipitation effects in the presence of hydroxide. Water heterolysis requires catalytic sites for the potential-driven dissociation reaction. Sharp edges at the 2D-titanate surfaces (causing high local field gradients) with a pK$_A$ close to 4 (resulting in active sites for field-driven water dissociation) could provide well-suited reaction environments, but further study will be needed to confirm this. A more quantitative observation of the relative contributions of proton and electrolyte cation transport to the diode current will be required. NBu$_4^+$ cations have been shown to suppress cation conduction and ultimately 'invert' the diode to anion conduction due to an increase in positive charge in the lamellar space. In order to improve the understanding of these processes, in future, theory could be developed at both the atomic level and the meso-level to take in account localised field effects as well as interfacial concentration gradient/polarisation effects that are caused by the current flow.

Applications of titanate films could be possible (i) under low ionic strength conditions as a catalyst for water heterolysis [55] or (ii) under higher ionic strength conditions in desalination. However, the ability of the lamellar structures to maintain effective rectification at higher ionic strengths is currently poor. Therefore, the inter-lamellar space needs to be better designed to maintain a higher rectification effect (by maintaining semi-permeability). This could be achieved by tuning the concentration and structure of the tetraalkylammonium guest cations. Further work will be directed toward the effects of inter-lamellar guests on the mechanism and the use of nanosheet materials in desalination and sensing.

**Acknowledgements** B.R.P. would like to thank the Indonesian Endowment (LPDP RI) for a PhD scholarship.